\documentclass[prl,twocolumn]{revtex4-1} % for arXiv
\usepackage[utf8]{inputenc}
\usepackage{graphicx} % for arXiv
\usepackage{amsmath} % for arXiv
\usepackage{amssymb} % for arXiv
\usepackage{multirow}
\usepackage{ulem}
\usepackage{hyperref}
\usepackage{siunitx}
\usepackage{color}

\hypersetup{
  pdfauthor={Florian Sedlmeir},
  pdftitle={Polarization-selective out-coupling of whispering gallery modes},
  pdfsubject={Physics, Optics, Whispering Gallery Mode Resonators, polarization, birefringence, coupling},
  urlcolor=blue,
}

\draft
\newcommand{\nii}[1]{_{\mathrm{#1}}}

\newcommand{\authoro}[2][]{\author{#2$^{#1}$}}    % for arXiv
\newcommand{\affil}[2][]{\affiliation{$^{#1}$#2}} % for arXiv

\begin{document}

\title{Polarization-selective out-coupling of whispering gallery modes}

 \authoro[1,2,3,4,*,+]{Florian Sedlmeir}
 \authoro[1,+]{Matthew R.\ Foreman}
 \authoro[1,2]{Ulrich Vogl}
 \authoro[1]{Richard Zeltner}
 \authoro[1,2,3]{Gerhard Schunk}
 \authoro[1]{Dmitry V.\ Strekalov}
 \authoro[1,2]{Christoph Marquardt}
 \authoro[1,2]{Gerd Leuchs}
 \authoro[4]{Harald G.\ L.\ Schwefel\vspace{2ex}}
 \affil[1]{Max Planck Institute for the Science of Light, Günther-Scharowsky-Straße 1/Building 24, 90158 Erlangen, Germany}
\affil[2]{Institute for Optics, Information and Photonics, University Erlangen-Nürnberg, Staudtstr.\ 7/B2, 91058 Erlangen, Germany}
\affil[3]{SAOT, School in Advanced Optical Technologies, Paul-Gordan-Str.\ 6, 91052 Erlangen, Germany}
\affil[4]{Department of Physics, University of Otago, Dunedin, New Zealand\vspace{1ex}}

\affil[*]{Corresponding author: florian.sedlmeir@web.de}

\affil[+]{these authors contributed equally to this work}

\date{\today}

\begin{abstract}
Whispering gallery mode (WGM) resonators are an important building block for linear, nonlinear and quantum optical experiments. In such experiments, independent control of coupling rates to different modes can lead to improved conversion efficiencies and greater flexibility in generation of non-classical states based on parametric down conversion. In this work, we introduce a scheme which enables selective out-coupling of WGMs belonging to a specific polarization family, while the  orthogonally polarized modes remain largely unperturbed. Our technique utilizes material birefringence in both the resonator and coupler such that a negative (positive) birefringence allows selective coupling to TE (TM) polarized WGMs. We formulate a new coupling condition suitable for describing the case where the refractive indices of the resonator and the coupler are almost the same, from which we derive the criterion for polarization-selective coupling. We experimentally demonstrate our proposed method using a lithium niobate disk resonator coupled to a lithium niobate prism, where we show a \SI{22}{dB} suppression of coupling to TM modes relative to TE modes.
\end{abstract}


\maketitle %\maketitle must follow title, authors, abstract and \pacs

Convex shaped dielectric resonators, such as disks and spheres, can support whispering gallery modes (WGMs) by continuous total internal reflection of light at their surface. WGMs possess a number of unique properties which makes them a natural choice for the study and exploitation of nonlinear optical effects \cite{breunig_three-wave_2016,strekalov_nonlinear_2016}. Firstly, since WGMs are guided by total internal reflection they are spatially well confined within the resonator. Together with extremely low losses, or equivalently, high quality ($Q$) factors, this leads to a strong intra-cavity field. Moreover, high $Q$ resonances can be supported over the complete spectral range for which the dielectric is transparent enabling efficient all-resonant frequency mixing of vastly different wavelengths. For instance, conversion from microwave to optical frequencies \cite{ilchenko_whispering-gallery-mode_2003,rueda_efficient_2016} has recently been demonstrated. Further examples include second harmonic generation (ranging from visible\cite{furst_naturally_2010} to ultraviolet wavelengths \cite{furst_second-harmonic_2015}) and resonantly enhanced parametric down conversion \cite{furst_low-threshold_2010,werner_blue-pumped_2012}. Octave spanning frequency combs have also been realized on a WGM platform \cite{delhaye_octave_2011}. 

WGM resonators are commonly used with external dielectric couplers, such as tapered fibres \cite{cai_observation_2000} or prisms \cite{braginsky_quality-factor_1989}, which are placed within the evanescent field of the WGM such that it is partially converted to a propagating wave in the coupler (or vice versa). Efficient coupling, however, requires phase-matching between the resonator and the coupler fields, that is to say the tangential projections of their wavevectors must be equal. Evanescent coupling nevertheless allows the coupling strength to be continuously tuned by adjusting the distance between the coupler and resonator. Control of the coupling rates is desirable since it allows the optimal experimental regime to be reached, as determined by the intrinsic losses and, in nonlinear experiments, nonlinear coupling coefficients \cite{sturman_generic_2011}. Several applications can also benefit greatly from the ability to independently tune the signal bandwidth without adversely affecting the orthogonally polarized pump mode, such as cavity assisted photon pair generation by spontaneous parametric down conversion \cite{Fortsch2013} where tunability allows individual atomic transitions to be addressed \cite{Schunk2015}.  Evanescent coupling strongly depends on the wavelength of the light, but only weakly on the mode polarization. Independent tuning of the coupling rates of transverse magnetic (TM) or transverse electric (TE) modes is therefore not possible with isotropic WGM resonators and couplers. Material birefringence has, however, been shown to facilitate polarization-selective coupling to modes of a dielectric ring resonator \cite{Fiedler1993}. Specifically, if either the prism, the resonator, or both are birefringent, the effective refractive index experienced by TE and TM modes differs. Accordingly, one can access a regime where phase-matching is fulfilled for only one polarization family.

In this article, we describe polarization-selective coupling using a prism and WGM resonator which are both fabricated from the same birefringent material. In this case, the refractive indices of the resonator and prism are comparable although may not be identical depending on the orientation of their respective optic axes. For this system, we formulate a selective coupling criterion which allows us to establish which birefringent crystals are suitable for realization of polarization-selective coupling. We find that the finite extent of the coupling region (or `coupling window') necessitates modification of the commonly used coupling condition \cite{gorodetsky_high-q_1994}. Furthermore, we experimentally demonstrate selective coupling with a $z$-cut lithium niobate (LN) resonator (optic axis parallel to the symmetry axis) and an $x$-cut LN prism (optic axis perpendicular to that of the resonator). We find that  coupling of TM modes is suppressed by \SI{22}{dB} relative to that of heavily overcoupled TE modes. With critically coupled TE modes, coupling to TM modes was not measurable.

\begin{figure}
\centering
\includegraphics[clip,angle=0,width=\linewidth]{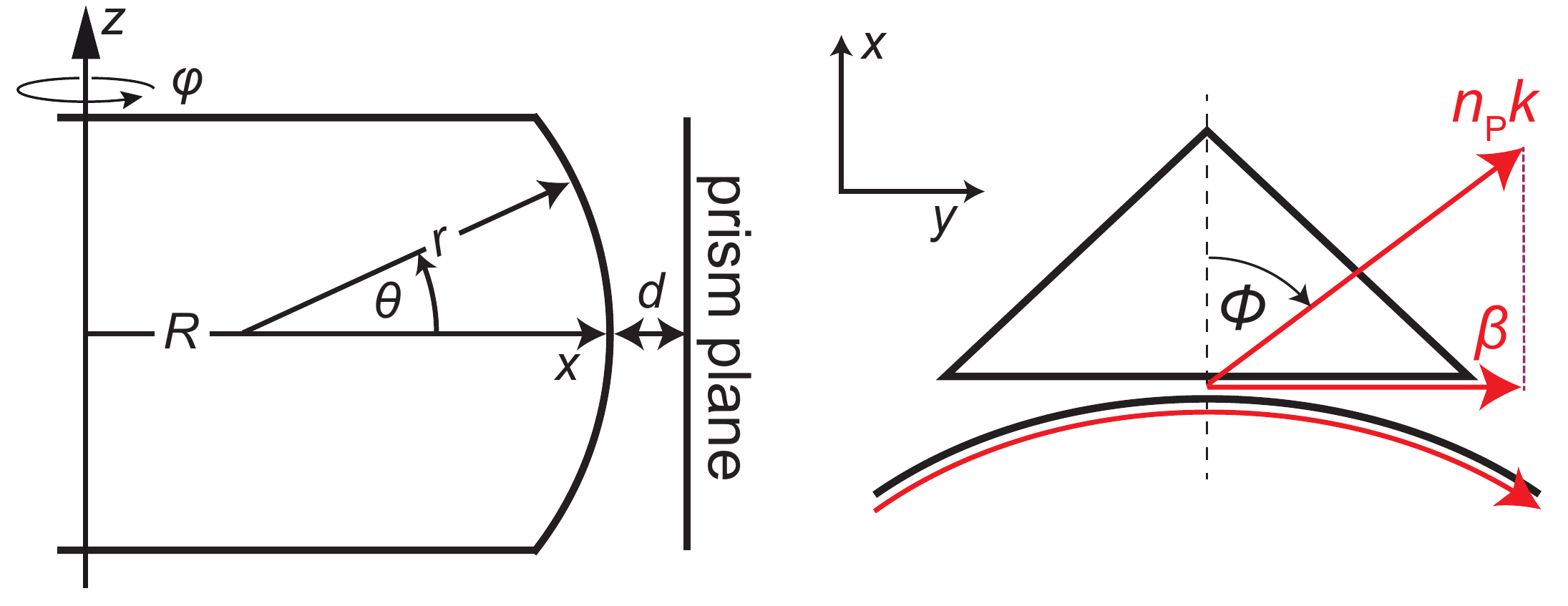}
\caption[]{(Left) Coordinate system and geometrical parameters considered in this work. (Right) Illustration of the coupling condition within the ray optics picture.}
\label{fig:framework}
\end{figure}

To first order, evanescent coupling of a WGM to a prism can be described using ray optics, whereby the WGM is treated as a plane wave propagating at a near glancing angle to the resonator surface. The tangential component of the wavevector is equal to the propagation constant, $\beta = k \bar{n}\nii{r}$, of the WGM, where $k$ is the wavenumber in vacuum and $\bar{n}\nii{r}$ is the geometry dependent effective refractive index of the mode \cite{strekalov_nonlinear_2016}. Application of Snell's law then shows that the emission angle of the WGM into the prism is $\phi \approx  \arcsin( \bar{n}\nii{r} / n\nii{p})$, where $n\nii{p}$ denotes the refractive index of the prism (see Fig.~\ref{fig:framework}). Coupling can therefore only be achieved if $n\nii{p} \geq \bar{n}\nii{r}$. If this condition is not satisfied, only evanescent waves are formed in the prism and no energy is carried away from the resonator. 

Closer examination of the coupling process reveals that the actual coupling condition is less strict than that given by geometric optics. Given that the resonator surface is curved, and that the evanescent field outside the resonator decays quickly with distance, coupling occurs in a strongly localized region of space between the resonator and the coupler. This spatially confined field distribution can be described by an infinite sum of plane waves with their wavenumbers distributed around the propagation constant $\beta$ of the WGM. Consequently, even if the refractive index of the prism is slightly smaller than that of the resonator, coupling can still be achieved through a wing of this distribution. 

In order to derive the coupling condition more quantitatively, we consider the WGM field distribution on the exterior surface of a disk shaped resonator with major and minor radius of curvature $R$ and $r$ respectively (see Fig.~\ref{fig:framework}). The field on the surface of the resonator can be expressed as\cite{breunig_whispering_2013}
\begin{equation}
\Psi^{\text{r}}(\theta,\varphi) \propto H_p\!\left[\frac{\theta}{\theta_m}\right]  \exp\left[ -\frac{\theta^2}{2\theta_m^2}\right] \exp [i m\varphi],
\label{equ:mode_structure_spherical}
\end{equation}
where $\theta$ and $\varphi$ are the polar and the azimuthal angles shown in Fig.~\ref{fig:framework}, $H_p$ are the $p$-th order Hermite polynomials, $m$ and $p$ are the azimuthal and polar mode numbers and $\theta_m = (R/r)^{(3/4)}/\sqrt{m}$. 
Due to the evanescent nature of the mode outside the resonator, which can be well approximated by an exponential decay \cite{demchenko_analytical_2013}, the size of the coupling spot is small compared to the resonator size. We therefore can make the small angle approximations $\theta \approx z/r$ and $\varphi \approx y/R$.
The mode profile at the prism interface can furthermore be related to that at the resonator surface via $\Psi^{\text{p}} = T \Psi^{\text{r}}$, where
\begin{equation}
T (y,z)  = \exp\left[- \kappa \left(d + \frac{z^2}{2 r } + \frac{y^2}{2 R}\right)\right]
\label{equ:gaussian_coupling_window}
\end{equation}
describes the Gaussian coupling window \cite{gorodetsky_high-q_1994}. The decay constant is given by $\kappa \approx k \sqrt{\bar{n}\nii{r}^2-n\nii{h}^2}$, where $n\nii{h}$ is the refractive index of the medium surrounding the resonator and $d$ is the smallest distance between the prism and the resonator (see Fig.~\ref{fig:framework}). While the $y$ dependence of $\Psi^{\text{p}}$ is thus ultimately determined only by the size of the coupling window and the angular momentum $m$ of the WGM, the $z$ behaviour exhibits an additional dependence on the polar shape of the mode. The most commonly addressed WGM is the Gaussian shaped fundamental mode ($p=0$) for which $H_0 = 1$. In this case the field in the plane of the coupling prism reduces to
\begin{align}
\!\!\!\Psi^{\text{p}} (y,z)  \propto \exp\left[-\frac{1}{2}\left(\frac{y^2}{\Delta y^2}+\frac{z^2}{\Delta z^2} \right)\right]\exp \left[i\beta y -\kappa d\right],
\label{equ:xy_modes_gaussian_window_simplified}
\end{align}
where $\Delta y^2= R/{\kappa}$, $\Delta z^{-2} = (r\theta_m)^{-2}   + ({\kappa}/{r})^{-2}$ and the propagation constant is given by $\beta = m/R = k \bar{n}\nii{r}$.
 The field distribution at the prism surface can now be analytically decomposed into its Fourier components yielding\cite{foreman_dielectric_2016}
\begin{equation}
\widetilde{\Psi}^\text{p}(k_y,k_z) \propto \exp\left[-\frac{\Delta y^2}{2}\left(k_y - \beta \right)^2 + \frac{\Delta z^2}{2} k_z \right].
\end{equation}
The spatial Fourier components in the $k_y$ direction are distributed around $\beta$ with a spread of $\Delta k_y = 1/\Delta y$, whereas the $k_z$ spectrum is centered on zero with a width of $\Delta k_z = 1/\Delta z$. Similarly to above, plane wave components with $k_y^2 + k_z^2 \leq n\nii{p}^2 k^2$ generate travelling waves upon transmission at the prism and therefore couple out of the resonator. Noting that $\Delta k_z \ll n_p k$, we can however safely neglect the $k_z$ dependence. Designating a WGM to be coupled if $\beta - \Delta k_y \leq n\nii{p} k$, we hence obtain the modified coupling condition:
\begin{equation}
(\bar{n}\nii{r} - n\nii{p})k \leq \sqrt{\frac{\kappa}{R}} \quad \text{or} \quad \bar{n}\nii{r} - n\nii{p} \leq \sqrt{\frac{\sqrt{\bar{n}\nii{r}^2 - n\nii{h}^2}}{2 \pi R / \lambda}},
\label{equ:coupling_condition}
\end{equation}
where we have used $k = 2\pi/\lambda$. Consequently, coupling is possible even if $\bar{n}\nii{r} > n\nii{p}$ due to the finite extent of the coupling window. Note that the effective refractive index $\bar{n}\nii{r}$ is always smaller than the bulk index of the resonator but approaches it for sufficiently large resonators.
The deviation from the ray optics coupling condition, where $\Delta k_y = 0$, becomes greater for smaller resonators and longer wavelengths since the corresponding smaller coupling window leads to a larger $k$-spread.

Equation~\ref{equ:coupling_condition} also holds for the case of birefringent resonators and prisms. In this case, the polarization of the WGM determines the refractive index of the resonator and the prism. With a proper choice of material (i.e. magnitude of birefringence) and resonator size, selective coupling of WGMs belonging to a given polarization family can be achieved, while the orthogonally polarized modes remain uncoupled.
\begin{figure}
\centering
\includegraphics[clip,angle=0,width=\linewidth]{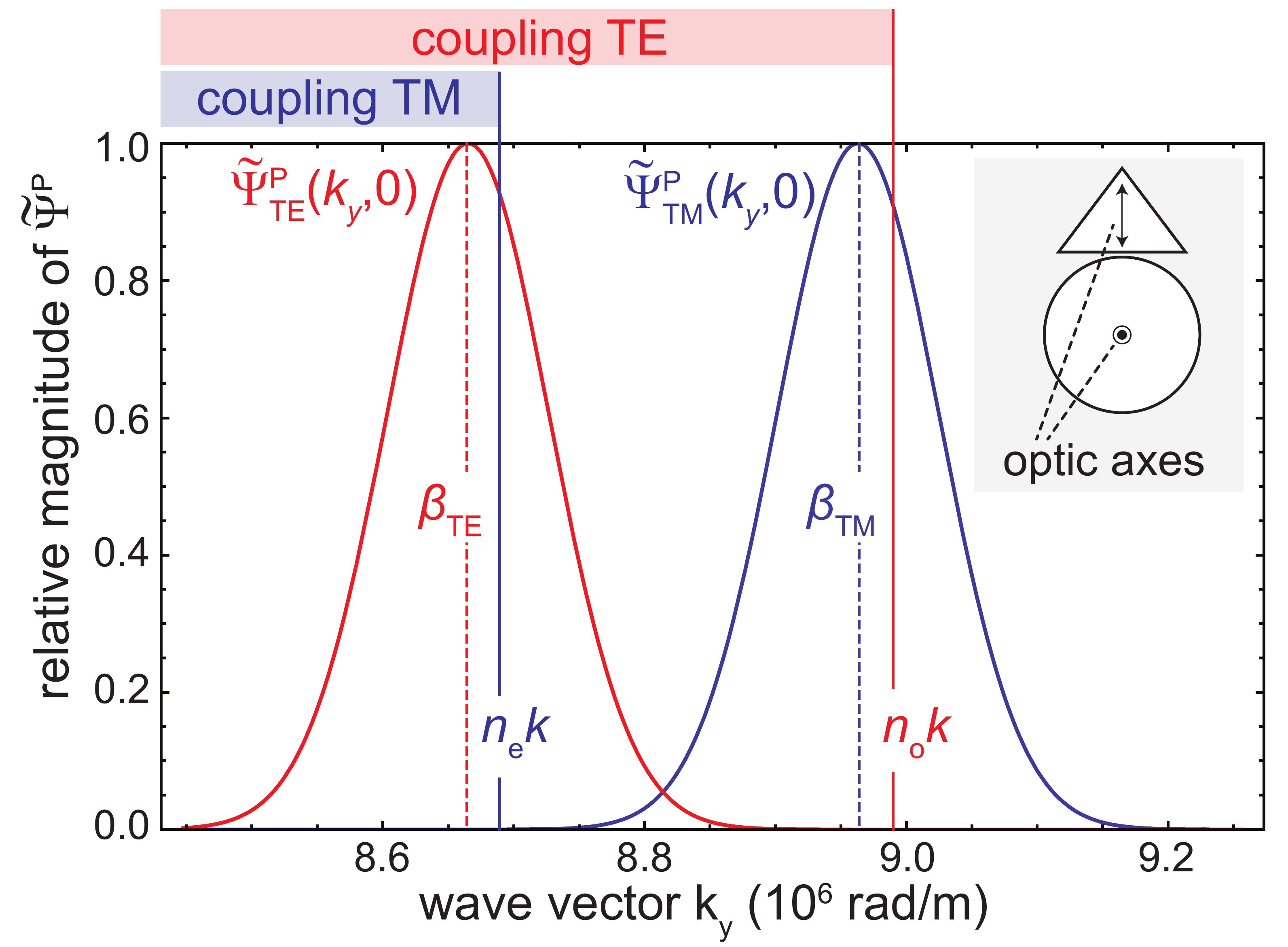}
\caption[]{The inset schematic shows the configuration considered in this work: the optic axis of the $z$-cut WGM resonator is perpendicular to the optic axis of the prism which are both made from the same material. The main plot shows the $k_y$ spectrum of a fundamental TE and TM mode at the prism interface for the case of a LN resonator ($\lambda =1550$~nm). The corresponding propagation constants and bulk wavenumbers inside the prism ($n\nii{o} k$ and $n\nii{e} k$) are also indicated. Note that coupling of the out-of-plane TE polarized mode (red) is possible, whereas it is not for the in-plane TM case (blue).}
\label{fig:lithium_niobate}
\end{figure}
Henceforth, we restrict attention to the case where both the resonator and prism are made from the same birefringent material, albeit with differently oriented optic axes. The basic configuration, shown in the inset of Fig.~\ref{fig:lithium_niobate}, is such that the optic axis of the resonator is aligned parallel to its symmetry axis ($z$-cut), whereas the optic axis of the prism is normal to the coupling plane ($x$-cut). Consequently, within the resonator the TE and TM modes experience the extraordinary ($n\nii{e}$) and ordinary ($n\nii{o}$) refractive index respectively. In the prism the TE mode is ordinarily polarized and sees $n\nii{o}$. The TM mode is extraordinarily polarized and therefore each constituent plane wave has its own angle dependent refractive index. Those waves propagating perpendicular to the optic axis, $k_x = 0$, experience a refractive index of $n\nii{e}$. Since this case corresponds to the transition from a propagating to an evanescent wave we take $n\nii{p} = n\nii{e}$ in Eq.~\eqref{equ:coupling_condition} when considering TM modes. Selective coupling can thus be achieved with this arrangement for a negatively birefringent crystal when
\begin{equation}
\left[\bar{n}\nii{r}(n\nii{o}) - n\nii{e}\right] k > \sqrt{\frac{\kappa(n\nii{o})}{R}}.
\label{equ:selective_coupling_condition}
\end{equation}
For a positive crystal, the ordinary and extraordinary refractive indices in Eq.~\eqref{equ:selective_coupling_condition} must be interchanged.
\begin{table}
\begin{tabular}{l|ccccc}
material & $R_{\text{min}}$ (mm) &$\Delta n$& $n_{\text{o}}$ & $n_{\text{e}}$ \\ \hline
lithium niobate \cite{Zelmon1997} &0.24 &$-0.073$&  2.211 & 2.138 & \\
barium borate \cite{zhang_optical_2000}& 0.09 & $-0.116 $& 1.647 & 1.531 \\
lithium tetraborate \cite{sugawara_linear_1998} & 0.31 & $-0.053$ & 1.589 & 1.536 \\
rutile \cite{devore_refractive_1951}& 0.04 & $\phantom{+}0.256$ &  2.453 & 2.709\\
crystal quartz \cite{ghosh_dispersion-equation_1999} & 8.41 & $\phantom{+}0.008$& 1.528 & 1.536 \\
magnesium fluoride \cite{Dodge1984} & 4.09 &$ \phantom{+}0.011$& 1.371 & 1.382\\
sapphire \cite{jeppesen_optical_1958} & 9.85 &$-0.008$ & 1.746 & 1.738 \\
\end{tabular}
\caption{Birefringent materials for which selective coupling using the scheme shown in Fig~\ref{fig:lithium_niobate} can be achieved. Note that the minimum resonator radius $R_{\text{min}}$ required for selective coupling, as determined from Eq.~(\ref{equ:selective_coupling_condition}), depends on wavelength (we assume $\lambda = \SI{1550}{nm}$), birefringence ($\Delta n = n\nii{o} - n\nii{e}$) and the effective refractive index. The latter was calculated using the dispersion relation given in \cite{breunig_whispering_2013} assuming $r=R/7$.\label{tab:materials}}
\end{table}
\begin{figure*}
\centering
\includegraphics[clip,angle=0,width=1\linewidth]{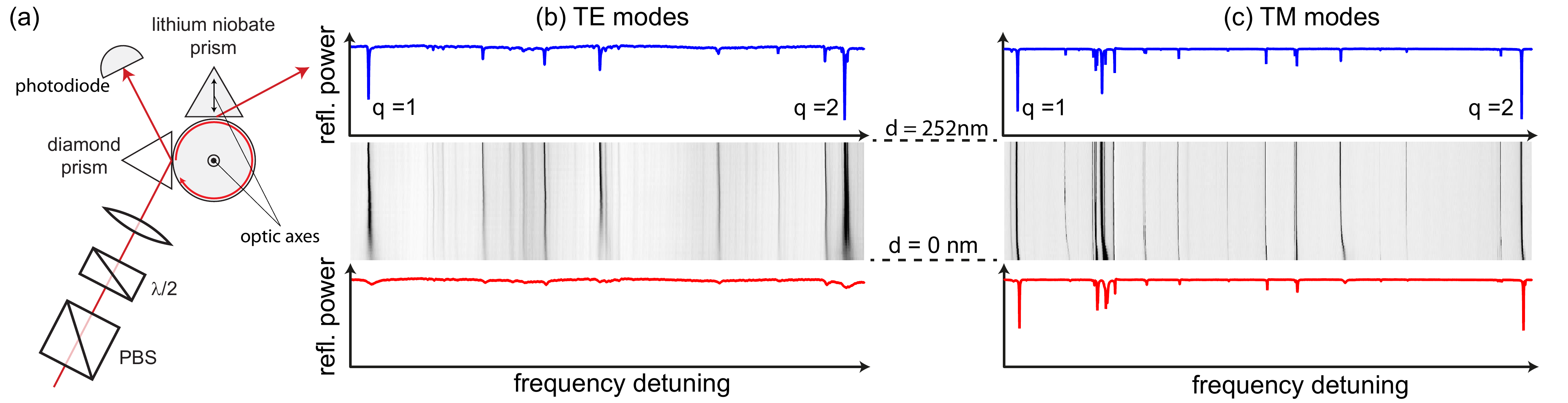}
\caption[]{(a) Schematic of the experimental setup. The grey intensity maps in panels (b) and (c) show the evolution of the spectra for the TE and TM modes, respectively, as the LN prism is brought closer to the resonator. The top (blue) spectra correspond to the critically coupled, almost unperturbed resonator (the prism-resonator distance is approximately \SI{252}{nm}), while the lower (red) plots show the spectra when the LN prism and resonator are in contact. }
\label{fig:selective_coupling_setup}
\end{figure*}

Table~\ref{tab:materials} lists a selection of birefringent resonator materials which can support high $Q$ WGMs and have been used, or have a potential use, in nonlinear optics. Also shown is the minimum resonator radius, $R_{\text{min}}$, required to achieve selective coupling of fundamental WGMs in a resonator in air, assuming an excitation wavelength of \SI{1550}{nm}. A strong dependence of $R_{\text{min}}$ on the magnitude of the birefringence $\Delta n = n_e - n_o$ is evident. A size limitation arises since for smaller resonators the coupling window narrows (broadening the corresponding angular spectrum) and  geometrical dispersion reduces the effective refractive index $\bar{n}\nii{r}$. For shorter wavelengths the condition on $R$ typically becomes less stringent due to decreased geometrical dispersion. Variation of $R_{\text{min}}$ with the minor radius $r$ is relatively weak. We note that Eq.~\eqref{equ:selective_coupling_condition} applies to polarization-selective \textit{out-coupling} of WGMs. Selective excitation (\textit{in-coupling}) of modes with differing polarization can easily be realised using an appropriately polarized pump beam. When the material birefringence is large, a proper choice of the in-coupling angle (when using prism coupling) can also lead to selective in-coupling.

In the remainder of this work we present an experimental demonstration of selective coupling of a $z$-cut LN resonator to an $x$-cut LN prism.
Since LN has a negative birefringence ($n\nii{o} > n\nii{e}$), TE modes can couple to the prism whereas coupling of TM modes is suppressed. We note that the converse holds for a material with positive birefringence. For  wavelengths $\lambda \approx \SI{1550}{nm}$ selective coupling is easily achieved for millimeter-sized LN resonators in air ($\Delta n=-0.073$, $\kappa \approx \SI{8}{\um^{-1}}$), as further illustrated in Figure~\ref{fig:lithium_niobate}, which shows the $k_y$ Fourier spectra for fundamental TE and TM modes in a LN resonator with $R = \SI{2.1}{mm}$ and $r = \SI{0.3}{mm}$. Practically all Fourier components of the TE mode ($\widetilde{\Psi}^\text{p}\nii{TE}(k_y,0)$, plotted in red) are smaller than the wavevector within the prism ($n\nii{o}k$) and consequently, coupling is allowed. In contrast, all components of the TM mode ($\widetilde{\Psi}\nii{TM}^\text{p}(k_{y},0)$, plotted in blue) are larger than the associated wavevector in the prism ($n\nii{e}k$). In this case, no coupling can occur. Each Fourier spectrum is peaked at the respective propagation constant $\beta$, which, by virtue of the geometric dispersion of WGMs, is slightly smaller than the associated wavevector of a plane wave in a bulk crystal. This reflects the fact that the effective refractive index, $\bar{n}\nii{r}$, is slightly smaller than the bulk index.

To experimentally demonstrate selective coupling, we used a two-port configuration comprising of an isotropic (non-selective) in-coupling prism and an anisotropic (selective) out-coupling prism  as shown in Fig.~\ref{fig:selective_coupling_setup}(a). A diamond turned $z$-cut LN resonator ($R = \SI{2.1}{mm}$, $r = \SI{0.3}{mm}$) was mounted onto a rotation stage. A narrowband, frequency tunable telecom laser ($\lambda \approx \SI{1550}{nm}$) was coupled into the resonator via a diamond prism after passing through a polarization controller. To determine the intrinsic loss rate of the resonator from the linewidths of the modes, the in-coupling prism was critically coupled to the WGM resonator. For the selective coupler we used a home-made $x$-cut LN prism. Both prisms were mounted on computer controlled piezo stages for precise distance control. The spectral positions and linewidths of excited modes were monitored in reflection after the in-coupling prism with a photodiode, as the LN prism was brought closer to the resonator. When light is coupled out of a given WGM through the LN prism, the linewidth and  coupling depth of the mode are modified since the out-coupling acts as an additional loss channel. The prisms were mounted orthogonally to each other to prevent the LN prism pushing the resonator towards the diamond prism, which would produce a change of the in-coupling rate hence masking the effect of the out-coupler.

The uppermost (blue) plots in Figs.~\ref{fig:selective_coupling_setup}(b) and (c) show the observed TE and TM mode spectra under critical coupling conditions, when the LN prism is far from the resonator. Since the in-coupling angle required for excitation of TE and TM modes differs, the two polarizations were measured separately. We determined the radial mode order $q$ via ordering the modes according to their free spectral range \cite{li_sideband_2012}, whereas the $p = 0$ modes were identified by inspection of their relative coupling contrast\cite{schunk_identifying_2014}. The loaded linewidth of the fundamental TE (TM) mode was measured to be $\Delta \nu\nii{TE} \approx \SI{2.2}{MHz}$ ($\Delta \nu\nii{TM} \approx \SI{1.0}{MHz}$) corresponding to a quality factor of $Q\nii{TE} \approx \SI{8.8 e 7}{}$ ($Q\nii{TM} \approx \SI{1.9 e 8}{}$). By increasing the voltage applied to the piezo in increments of \SI{0.1}{V}, the LN prism was slowly shifted towards the resonator in steps of approximately \SI{6.3}{nm} (see below). For each step, the reflected spectra are recorded and are shown in the grey-scale intensity maps of Figures~\ref{fig:selective_coupling_setup}(b) and (c). From top to bottom, the intensity map represents spectra for a prism-resonator separation varying from \SI{252}{nm} to \SI{0}{nm}. Both, TE and TM modes remain unperturbed until the prism draws close to the resonator ($d\sim 1/\kappa$), where the modes start to shift in frequency due to the dielectric interaction with the prism  \cite{foreman_dielectric_2016}. The TE modes show significant linewidth broadening and become strongly under-coupled as shown in the lower spectrum of Fig.~\ref{fig:selective_coupling_setup}(c). In contrast, the TM modes show almost no linewidth change as expected.

\begin{figure}
\centering
\includegraphics[clip,angle=0,width=1\linewidth]{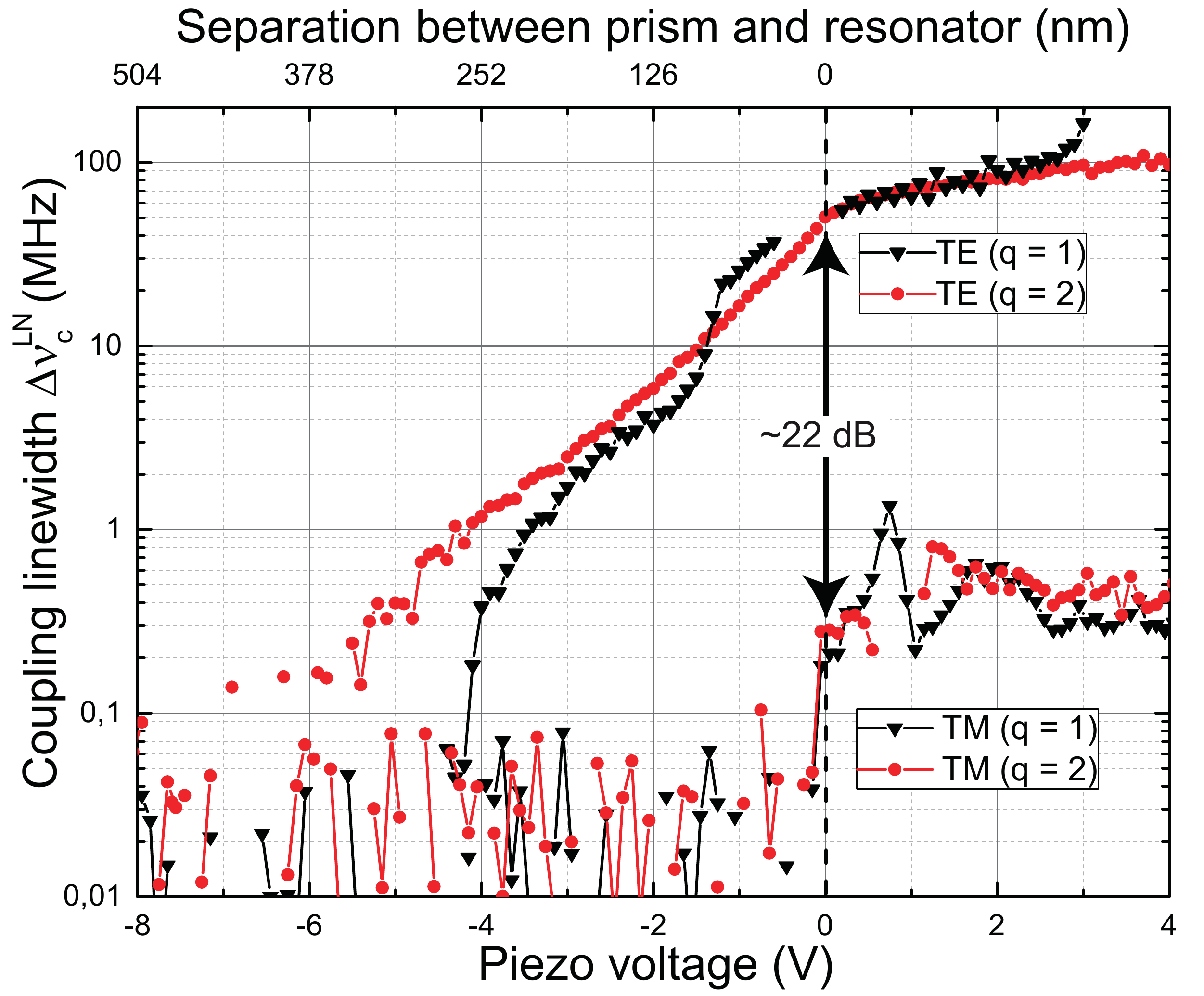}
\caption[]{WGM linewidth change induced by coupling to the LN prism for two different radial order TE and TM modes. The linewidth changes are plotted against the voltage applied to the piezo (bottom axis) and the corresponding estimated prism displacement (top axis). The qualitative change of the linewidths is used to identify the mechanical contact point between the prism and the resonator. The linewidth of the TE modes increases by approximately 53\,MHz before contact, while the linewidth of the TM modes is largely unchanged.}
\label{fig:linewidth_TE_TM}
\end{figure}
For quantification, we  selected the two lowest radial order modes ($q = 1, 2$) for both TE and TM polarizations and determined the associated resonance linewidths for every voltage step by means of Lorentzian fitting. The measured linewidths $\Delta \nu$ can be expressed as $\Delta \nu\nii{c} + \Delta \nu\nii{i} + \Delta \nu\nii{c}^\text{LN}$, where $\Delta \nu\nii{c}$ represents the constant coupling rate to the diamond prism, $\Delta \nu\nii{i}$ denotes the intrinsic loss rate and $\Delta \nu\nii{c}^\text{LN}$ is the variable coupling rate to the LN prism. Subtraction of the unperturbed linewidth (measured when the LN prism is far from the resonator) from those determined as the LN prism is brought closer allows $\Delta \nu\nii{c}^\text{LN}$ to be found. The experimentally determined coupling rates, $\Delta \nu\nii{c}^\text{LN}$, are plotted in Fig.~\ref{fig:linewidth_TE_TM} as a function of the relative voltage applied to the piezo shifting the LN prism. An estimate of the distance travelled by the piezo per volt can also be deduced from the variation of the TE coupling rate by noting that $\Delta \nu\nii{c}^{\text{LN}} \propto e^{- 2 \kappa d}$, which yields a piezo displacement of approximately \SI{63}{nm/V}. We show the estimated prism-resonator separation on the upper axis of Figure~\ref{fig:linewidth_TE_TM}. 
When plotted on a logarithmic scale, the coupling linewidth increases with a constant gradient as $d$ is decreased. The strong change in the gradient, seen in Figure~\ref{fig:linewidth_TE_TM}, was used to determine the position at which the prism and resonator come into mechanical contact.
All voltages and distances have been defined relative to this point. At the  contact point, the  coupling rate of TE modes to the LN prism reaches \SI{53}{MHz}, whereas the TM modes only weakly couple at a rate of \SI{0.3}{MHz}. Out-coupling of the TM modes relative to the TE modes is therefore suppressed by a factor of 177, or equivalently about $\SI{22}{dB}$. Further increase of the piezo voltage pushes the LN prism into the resonator leading to further increases in the observed linewidths, attributable to deformation of the resonator and an associated higher intrinsic loss. The step-like increase in the TM linewidth upon contact is ascribed to these additional losses or alternatively to a slight modification of the distance between the in-coupling diamond prism and the resonator.

In conclusion, we have derived a criterion for polarization-selective coupling using a prism and resonator of the same material. Many commonly used nonlinear crystals fulfil this condition. We have shown that an $x$-cut LN prism can be used to tune the rate at which TE modes of a $z$-cut LN resonator are out-coupled without disturbing the bandwidth of the TM modes. This provides a useful tool for nonlinear optical experiments in birefringent WGM resonators allowing for independent control of coupling to differently polarized modes, which in many schemes correspond to the pump and the signal modes. We also provide a generalized coupling theory which accounts for the finite spatial extent of the coupling window. As a consequence, we found that coupling to WGMs is  possible even if the refractive index of the prism is slightly smaller than the WGM effective index.

%\bibliography{selective_coupling}

\begin{thebibliography}{10}
	
	\bibitem{breunig_three-wave_2016}
	I.~Breunig, Laser Phot. Rev. \textbf{10}, 569 (2016).
	
	\bibitem{strekalov_nonlinear_2016}
	D.~V. Strekalov, C.~Marquardt, A.~B. Matsko, H.~G.~L. Schwefel, and G.~Leuchs,
    arXiv:1605.07972 (2016). 
	
	\bibitem{ilchenko_whispering-gallery-mode_2003}
	V.~S. Ilchenko, A.~A. Savchenkov, A.~B. Matsko, and L.~Maleki,
	J. Opt. Soc. Am. B \textbf{20},	333 (2003).
	
	\bibitem{rueda_efficient_2016}
	A.~Rueda, F.~Sedlmeir, M.~C. Collodo, U.~Vogl, B.~Stiller, G.~Schunk, D.~V.
	Strekalov, C.~Marquardt, J.~M. Fink, O.~Painter, G.~Leuchs, and H.~G.~L.
	Schwefel, Optica \textbf{3}, 597 (2016).
	
	\bibitem{furst_naturally_2010}
	J.~U. F\"{u}rst, D.~V. Strekalov, D.~Elser, M.~Lassen, U.~L. Andersen,
	C.~Marquardt, and G.~Leuchs, Phys. Rev. Lett. \textbf{104}, 153901 (2010).
	
	\bibitem{furst_second-harmonic_2015}
	J.~U. F\"{u}rst, K.~Buse, I.~Breunig, P.~Becker, J.~Liebertz, and L.~Bohat\'{y},
	 Opt. Lett. \textbf{40}, 1932 (2015).
	
	\bibitem{furst_low-threshold_2010}
	J.~U. F\"{u}rst, D.~V. Strekalov, D.~Elser, A.~Aiello, U.~L. Andersen,
	C.~Marquardt, and G.~Leuchs, Phys. Rev. Lett. \textbf{105}, 263904 (2010).
	
	\bibitem{werner_blue-pumped_2012}
	C.~S. Werner, T.~Beckmann, K.~Buse, and I.~Breunig, Opt. Lett.
	\textbf{37}, 4224 (2012).
	
	\bibitem{delhaye_octave_2011}
	P.~Del’Haye, T.~Herr, E.~Gavartin, M.~L. Gorodetsky, R.~Holzwarth, and T.~J.
	Kippenberg, Phys. Rev. Lett. \textbf{107}, 063901 (2011).
	
	\bibitem{cai_observation_2000}
	M.~Cai, O.~Painter, and K.~J. Vahala, Phys. Rev. Lett. \textbf{85},
	74 (2000).
	
	\bibitem{braginsky_quality-factor_1989}
	V.~B. Braginsky, M.~L. Gorodetsky, and V.~S. Ilchenko, Phys. Lett. A \textbf{137}, 393 (1989).
	
	\bibitem{sturman_generic_2011}
	B.~Sturman and I.~Breunig, J. Opt. Soc. Am. B \textbf{28}, 2465 (2011).
	
	\bibitem{Fortsch2013}
	M.~F{\"{o}}rtsch, J.~F{\"{u}}rst, C.~Wittmann, D.~Strekalov, A.~Aiello, M.~V.
	Chekhova, C.~Silberhorn, G.~Leuchs, and C.~Marquardt, Nat. Commun. \textbf{4}, 1818 (2013).
	
	\bibitem{Schunk2015}
	G.~Schunk, U.~Vogl, D.~V. Strekalov, M.~F{\"{o}}rtsch, F.~Sedlmeir, H.~G.~L.
	Schwefel, M.~G{\"{o}}belt, S.~Christiansen, G.~Leuchs, and C.~Marquardt,
	 Optica \textbf{2}, 773	(2015).
	
	\bibitem{Fiedler1993}
	K.~Fiedler, S.~Schiller, R.~Paschotta, P.~K{\"u}rz, and J.~Mlynek,
	 Opt. Lett.	\textbf{18}, 1786 (1993).
	
	\bibitem{gorodetsky_high-q_1994}
	M.~Gorodetsky and V.~Ilchenko, Opt. Commun. \textbf{113}, 133 (1994).
	
	\bibitem{breunig_whispering_2013}
	I.~Breunig, B.~Sturman, F.~Sedlmeir, H.~G.~L. Schwefel, and K.~Buse,
	Opt. Express \textbf{21}, 30683 (2013).
	
	\bibitem{demchenko_analytical_2013}
	Y.~A. Demchenko and M.~L. Gorodetsky, J. Opt. Soc. Am. B \textbf{30},
	3056 (2013).
	
	\bibitem{foreman_dielectric_2016}
	M.~R. Foreman, F.~Sedlmeir, H.~G.~L. Schwefel, and G.~Leuchs, arXiv:1607.05098 (2016).
	
	\bibitem{Zelmon1997}
	D.~E. Zelmon, D.~L. Small, and D.~Jundt, J. Opt. Soc. Am. B
	\textbf{14}, 3319 (1997).
	
	\bibitem{zhang_optical_2000}
	D.~Zhang, Y.~Kong, and J.-Y. Zhang, Opt. Commun. \textbf{184}, 485 (2000).
	
	\bibitem{sugawara_linear_1998}
	T.~Sugawara, R.~Komatsu, and S.~Uda, Solid State Commun. \textbf{107}, 233 (1998).
	
	\bibitem{devore_refractive_1951}
	J.~R. Devore, J. Opt. Soc. Am. \textbf{41}, 416 (1951).
	
	\bibitem{ghosh_dispersion-equation_1999}
	G.~Ghosh, Opt. Commun.
	\textbf{163}, 95--102 (1999).
	
	\bibitem{Dodge1984}
	M.~J. Dodge, Appl. Opt. \textbf{23}, 1980--1985 (1984).
	
	\bibitem{jeppesen_optical_1958}
	M.~A. Jeppesen, J. Opt. Soc. Am. \textbf{48}, 629 (1958).
	
	\bibitem{li_sideband_2012}
	J.~Li, H.~Lee, K.~Y. Yang, and K.~J. Vahala, Opt. Express \textbf{20},
	26337--26344 (2012).
	
	\bibitem{schunk_identifying_2014}
	G.~Schunk, J.~U. F\"urst, M.~F\"ortsch, D.~V. Strekalov, U.~Vogl, F.~Sedlmeir,
	H.~G.~L. Schwefel, G.~Leuchs, and C.~Marquardt, Opt. Express \textbf{22}, 30795--30806 (2014).

	\end{thebibliography}

\end{document}